\documentclass[pra,aps,twocolumn,nopacs,superscriptaddress,nofootinbib]{revtex4}
\usepackage{graphicx}
\usepackage{dcolumn}
\usepackage{bm}
\usepackage{amsmath}
\usepackage{epsfig}

\begin{document}
\title{Quantum Junction Plasmons in Graphene Dimers}
\author{Sukosin~Thongrattanasiri}
\affiliation{IQFR - CSIC, Serrano 119, 28006 Madrid, Spain}
\author{Alejandro~Manjavacas}
\affiliation{IQFR - CSIC, Serrano 119, 28006 Madrid, Spain}
\author{Peter~Nordlander}
\affiliation{Dept. of Physics and Astronomy, M.S. 61, Rice University, Houston, Texas 77005-1892, USA}
\author{F.~Javier~Garc\'{\i}a~de~Abajo}
\email{J.G.deAbajo@nanophotonics.es}
\affiliation{IQFR - CSIC, Serrano 119, 28006 Madrid, Spain}

\begin{abstract}
The interaction between doped graphene nanoislands connected by narrow junctions constitutes an ideal testbed to probe quantum effects in plasmonic systems. Here, the interaction between graphene plasmons in neighboring nanoislands is predicted to be extremely sensitive to the size and shape of the junctions. The reported {\it ab initio} calculations reveal three different regimes of interaction: (1) for narrow bridges ($<4$ carbon-atom rows), the conductance of the junction is too low to allow electron transport and the optical response is dominated by a characteristic bonding dipolar dimer mode that also appears in a classical description; (2) for wider junctions (4-8 carbon rows), a strong charge polarization is induced across the junction, which gives rise to a novel {\it junction plasmon} that has no counterpart in a classical description; (3) for even wider junctions ($\ge8$ rows), their conductance is sufficiently large to allow charge transport between the two graphene islands, resulting in a pronounced charge-transfer plasmon, which can also be described classically. This work opens a new path for the investigation of intrinsic plasmon quantum effects.
\end{abstract}
\maketitle

\section{Introduction}

A major advantage of plasmons --the collective excitations of conduction electrons in metals-- is that they can produce high levels of light confinement and optical field enhancement \cite{LSB03}, which enable them to strongly interact with objects that have relatively low optical cross-sections, such as atoms and simple molecules in solid-state environments \cite{paper156}. Additionally, confined plasmons can display intense dipoles through which they couple to external radiation. Therefore, plasmons can act as intermediaries (antennas) that boost the interaction of light with molecules (e.g., in techniques as useful as surface-enhance Raman scattering (SERS) \cite{KWK97,XBK99} and surface-enhanced infrared resonant absorption (SEIRA) \cite{KLN08}, and also in promising applications to tumor removal \cite{NHH04}, drug delivery \cite{LSH11}, improved photovoltaics \cite{AP10}, and catalysis \cite{JHE08}). An interesting scenario is found when a plasmon is supported by a single molecular structure rather than an extended nanoparticle. Many molecules have been found to exhibit plasmons \cite{KHS00,STK02,YYG07,YG08,ZLN12}, including the C$_{60}$ molecule and carbon nanotubes \cite{SSO91,KC92}. Like other carbon allotropes, they display plasmon bands at energies around $\sim5\,$eV ($\pi$ plasmons) and $\sim15-20\,$eV ($\sigma$ plasmons). However, these are conventional plasmons, in the sense that they are not too sensitive to the charge state of the molecules, similar to those found in noble metal nanoparticles. Instead, we are here concerned with plasmons hosted by doped graphene molecular nanostructures, which in contrast to noble-metal plasmons, only occur when they are electrically charged. These plasmons emerge at lower energies for typical levels of doping.

Graphene has recently emerged as an attractive electrically tunable, optical material \cite{LHJ08} displaying strong infrared plasmons, the frequencies of which scale roughly as $|n|^{1/4}$ with the doping charge density $n$ \cite{WSS06,HD07,HMZ09,JBS09}.  Graphene plasmons provide unprecedented levels of light confinement and field enhancement \cite{paper176}, which are promising tools for accessing quantum-optics phenomena \cite{CSH06,DSF10,SSR10,paper173}. In this context, graphene offers the additional benefits of a robust structure and a simple and convenient electrical tunability for controlling quantum interactions \cite{paper184}. The electrical modulation of graphene-plasmon-related phenomena also suggests potential application in optical waveguiding and signal processing \cite{VE11,NGG11,paper181}, metamaterials \cite{paper182,NPA12}, quantum optics \cite{paper176}, and spectrometry \cite{JGH11}. Experimental evidence of electrostatic control over plasmonic features in the absorption spectra of graphene has been recently reported \cite{JGH11}, followed by the observation and spatial mapping of confined plasmons in this material \cite{CBA12,FRA12}. Additionaly, graphene has been used to extrinsically modulate the plasmons of neighboring metallic structures \cite{ECN12,GPN12}. These advances have sparked considerable interest into the unique optical behavior of nanostructured graphene.

\begin{figure*}
\begin{center}
\includegraphics[width=170mm,angle=0,clip]{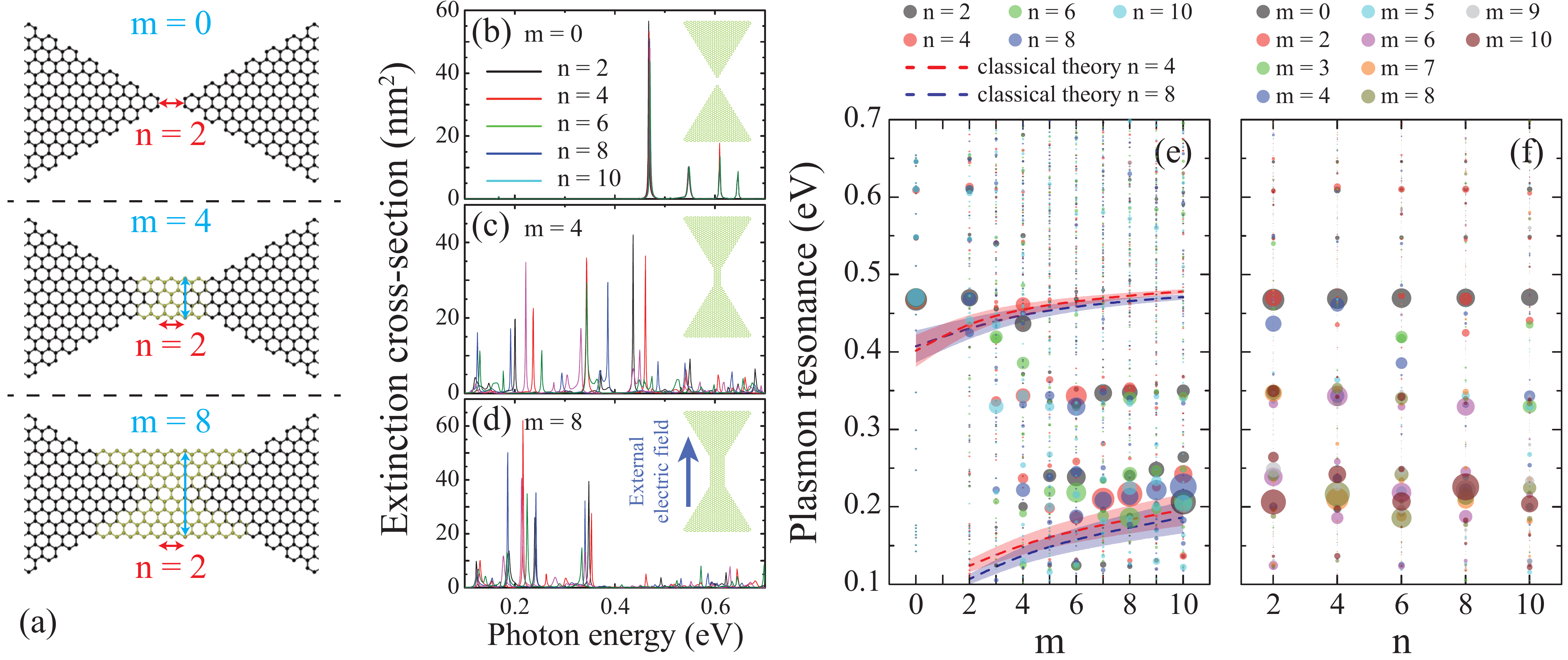}
\caption{{\bf Quantum junction plasmons in graphene bowties.} {\bf (a)} Details of the junction region in the structures under study, with definitions of bridge length $n$ and width $m$. {\bf (b-d)} Spectra for selected bridge widths $m=0,4,8$, and various lengths as indicated by different colors. The insets show the complete graphene structures for $n=2$. {\bf (e)} Plasmon resonances as a function of bridge width $m$. The color code for different lengths $n$ is given in the upper inset. The area of the circles is proportional to the area under the extinction peak for each plasmon feature. Classical-electrodynamics plasmon energies are shown by dashed curves for $n=4$ (red) and $n=8$ (blue), flanked by shaded areas that represent the strength of the modes. {\bf (f)} Plasmon resonances as a function of bridge length $n$. The graphene is doped to a Fermi energy $E_F=0.4\,$eV and has a mobility $\mu=10,000\,$cm$^2$/(V s). The full length of the bowties is 8\,nm.} \label{Fig1}
\end{center}
\end{figure*}

Here, we describe intrinsic quantum effects in the plasmonic response of bowtie graphene nanodimers bridged by thin junctions. Due to the two-dimensional character of this material, the addition of a small number of atoms ($<10$) is sufficient to dramatically modify the absorption spectrum of the entire dimer. Three distinct regimes of plasmonic behavior are predicted by our first-principles calculations. For separated nanotriangles bridged by narrow junctions ($<4$ atomic rows wide), we find a characteristic hybridized bonding dipolar plasmon (BDP) mode \cite{NOP04}. In the limit of wide junctions (more than 8 atomic rows), a dramatic reshaping of the optical spectrum leads to a pronounced charge-transfer plasmon (CTP), which emerges at lower energies, similar to what happens when the gap vanishes in a conventional metallic nanoparticle dimer \cite{paper075,LAH08,ZPN09,EBN12}. The plasmonic features in both of these regimes are also found within a classical electrodynamics description using the appropriate dielectric permittivity for the graphene, although the behavior of the CTP is substantially modified by quantum mechanical effects (see below). In contrast, for intermediate junction widths ($4-8$ atomic rows), a plasmonic feature shows up at intermediate energies, which is absent in classical-electrodynamics calculations. We refer to this mode as a junction plasmon (JP) because it is caused by quantum effects in the small {\it molecular} junction. We also predict a minor dependence on the length of the junction and a strong dependence on its width, thus confirming the importance of a quantum mechanical description of the optical properties of bridged graphene dimers.

\section{Results and discussion}
\label{results}

We simulate the optical response of nanoscale bowtie graphene structures using the random-phase approximation (RPA) and wave functions derived from a tight-binding description of the carbon sheet. Further details of the method are given elsewhere \cite{paper183}. The triangular structures forming the bowties are oriented with respect to the graphene lattice in such a way that they have armchair edges, which prevent undesired losses typically observed in zigzag-edge structures, and caused by the presence of zero-energy electronic edge states \cite{paper183}. Such states are not present in the structures under consideration, because they have the same number of atoms in both carbon sublattices \cite{FP07}. For comparison, we include classical-electrodynamics calculations, which are performed by describing the graphene through its local-RPA surface conductivity \cite{paper176} using a finite elements method (COMSOL). We fix the side length of the nanotriangles to 8\,nm in all cases ($\sim10^3$ atoms in each triangle). We set the Fermi energy of our structures to $E_F=0.4\,$eV, and the intrinsic damping to $\hbar\tau^{-1}=1.6\,$meV, corresponding to a DC mobility of 10,000\,cm$^2/($Vs$)$ \cite{JBS09}.

\begin{figure*}
\begin{center}
\includegraphics[width=170mm,angle=0,clip]{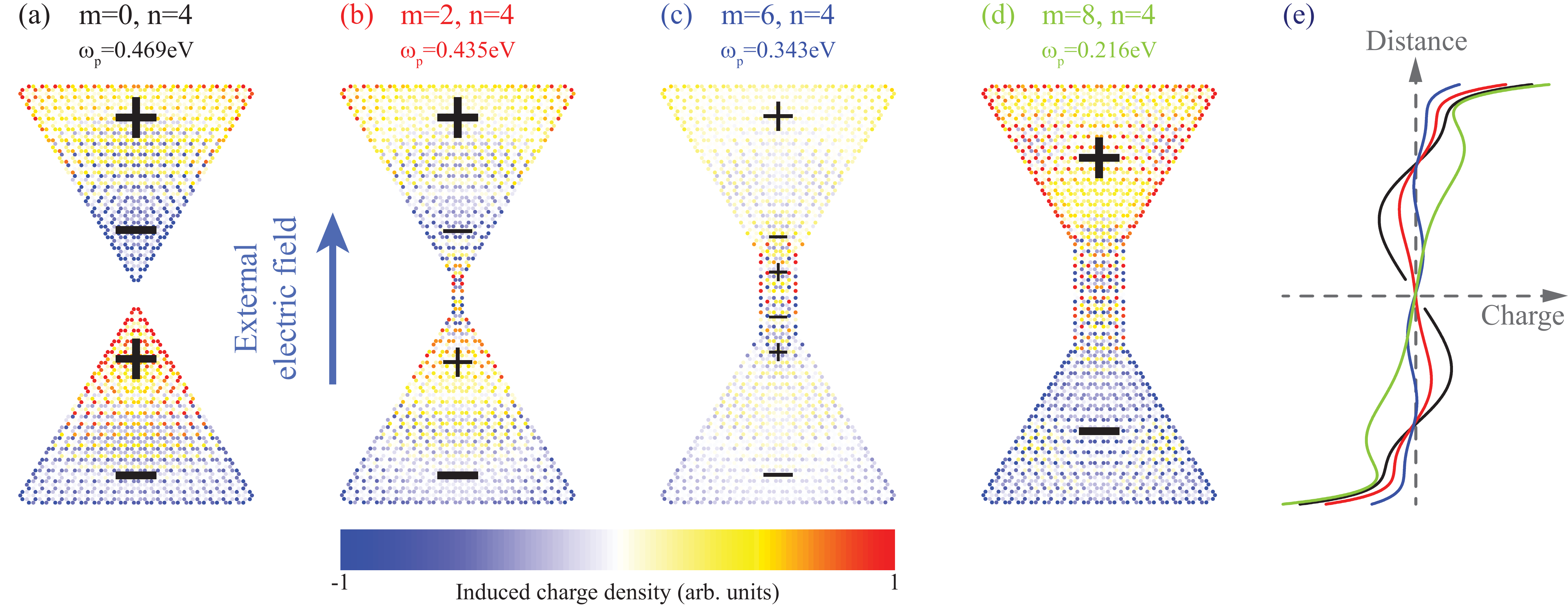}
\caption{{\bf Plasmon induced-charge distributions for representative graphene bowties.} {\bf (a-d)} Density plots with the color of each atom indicating its induced charge for different junction widths $m=0,2,6,8$ and the same length $n=4$. {\bf (e)} Induced charge integrated along the horizontal direction and averaged over four nearest carbon-atom neighbors.} \label{Fig2}
\end{center}
\end{figure*}

Our main results are summarized in Fig.\ \ref{Fig1}. We show details of the junction region in Fig.\ \ref{Fig1}(a) for three characteristic values of the width ($m=0,4,8$, where $m$ is the number of carbon-atom zigzag rows forming the junction, and $m=0$ stands for the non-touching configuration). The corresponding spectra are shown in Fig.\ \ref{Fig1}(b)-(d) for a range of junction lengths $n=2-10$ [$n$ is defined as the number of carbon hexagons that are needed to join the graphene triangles for $m=2$, see Fig.\ \ref{Fig1}(a)]. These spectra show a clear trend from high to low energy features as the junction width is increased, irrespective of the length. This behavior is further examined in Fig.\ \ref{Fig1}(e), where the plasmon features are arranged as a function of energy (vertical axis) and junction width ($m$, horizontal axis) for different values of the junction length $n$ (see color code in the upper inset; full spectra are provided in the Appendix (Fig.\ \ref{ext}). The area of the circles is made proportional to the intensity of the plasmon mode defined as the area under the corresponding plasmon peaks in the extinction spectra (for light polarization along the gap). For narrow bridges, the spectra exhibit prominent plasmonic features centered around $\sim0.47\,$eV. In contrast, the spectra for the wide-junction limit are dominated by lower-energy plasmonic features around $\sim0.22\,$eV. At intermediate values of the junction width, there is a complex transition between these two regimes, with intermediate-energy features around $\sim0.35\,$eV showing up in the spectra. The transition is fast, but not singular. These conclusions seem to be rather independent of the length of the junctions $n$. As Fig.\ \ref{Fig1}(e) illustrates, no significant spectral changes are induced when the plasmon energies are plotted as a function of junction length $n$. The three distinct behaviors of the  plasmon energies noted above for narrow, intermediate, and wide junctions appear for all values of the junction length.

Further insight into the character of these plasmons is provided by examining their induced-charge distributions, which are plotted in Fig.\ \ref{Fig2}(a)-(d) for representative structures along the transition from narrow to wide junction regimes. The polarization profile for the high-energy plasmon around $\sim0.47\,$eV in the non-touching dimer [$m=0$, Fig.\ \ref{Fig2}(a)] displays a dipole-dipole pattern that is not substantially modified when a narrow junction [$m=2$, Fig.\ \ref{Fig2}(b)] is present between the two triangular islands of the structure. This distinct polarization pattern is characteristic of a BDP. In contrast, the intermediate energy plasmon around $\sim0.35\,$eV [Fig.\ \ref{Fig1}(c)] for a junction width $m=6$ shows a clear dipolar polarization pattern across the junction. This is the JP that is formed from the local electronic properties of the junction. A weak charge transfer between the two nanotriangles is also evident in the plot and corresponds to the CTP, which is delocalized over the entire graphene structure like an standing wave. The latter shows a consistent $+-$ charge polarization that does not change with increasing $m$, as shown for $m=8$ in Fig.\ \ref{Fig2}(d). The CTP becomes the dominant feature once the junction is sufficiently wide. The signs superimposed on the density plots of Fig.\ \ref{Fig2}(a)-(d) qualitatively correspond to the distribution of the plasmon induced charge as a function distance to the dimer center, which is shown in Fig.\ \ref{Fig2}(e) (see also Fig.\ \ref{charge} in the Appendix). This analysis is rather independent of the length of the junction $n$. An in-depth inspection of bowties of increasing length leads to results that are consistent with the above conclusions, including the emergence of quantum junction plasmons once the width of the junction and the length of the structure are sufficiently large (see Fig.\ \ref{cut_end} in the Appendix for more details).

A classical-electrodynamics description of these graphene nanostructures (see broken curves in Fig.\ \ref{Fig1}(e) and Fig.\ \ref{classical} in the Appendix) does not completely reproduce the plasmonic behavior derived from first-principles. For instance, classical calculations miss the intermediate-energy junction plasmon that we observe in the quantum calculations for intermediate junction widths. Another discrepancy is that classical theory predicts a smooth fadeout of the dipole-dipole mode exhibited by non-overlapping triangles when the junction width is increased, while the CTP shows a singular behavior, as it migrates towards zero energy in the limit of vanishing junction width. This is in striking contrast to the results from the quantum description, in which the CTP is rather dispersionless and fades out smoothly in that limit [see Fig.\ \ref{Fig1}(e)]. The dipole-dipole and charge-transfer nature of the classically predicted plasmons is clear from their induced-charge distributions (see Appendix, Fig.\ \ref{classical}).

\begin{figure*}
\begin{center}
\includegraphics[width=160mm,angle=0,clip]{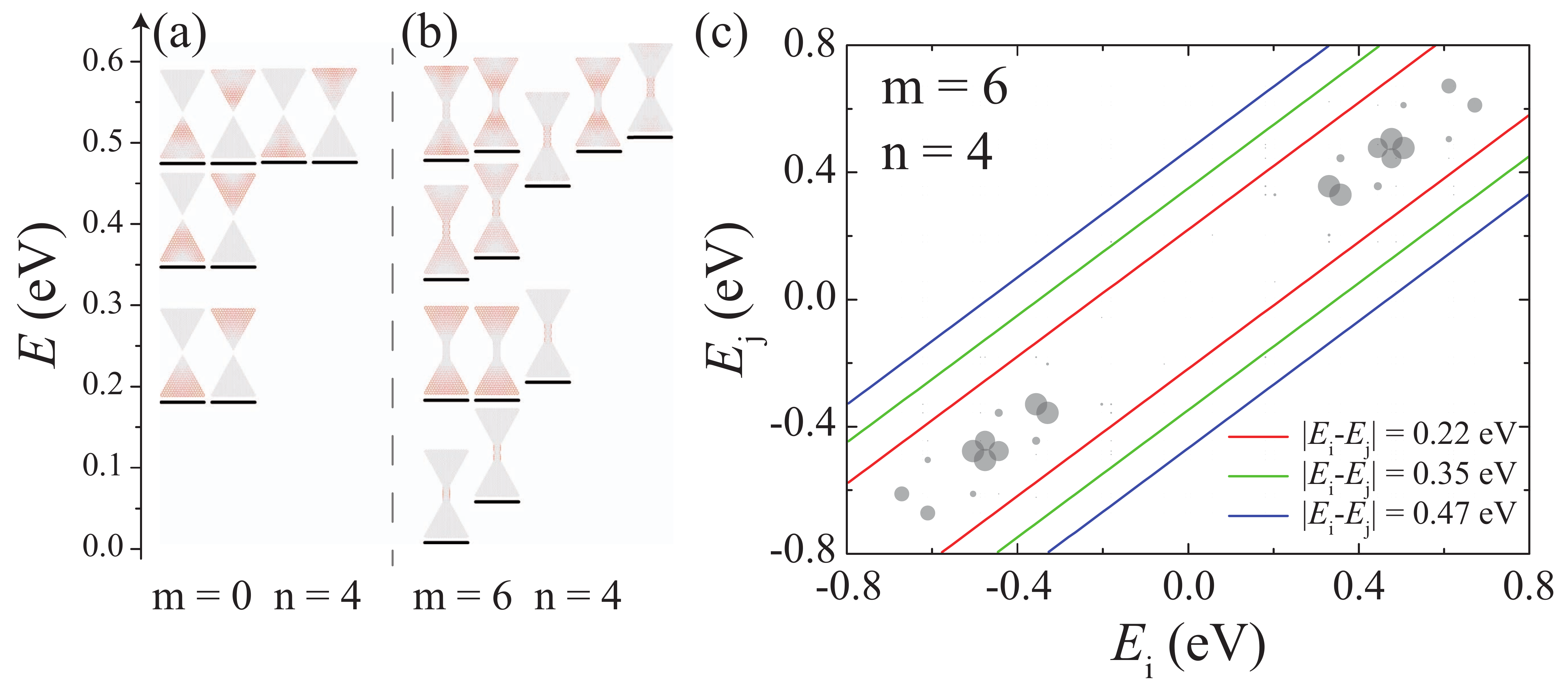}
\caption{{\bf Electronic junction states and electron-hole transition strengths.} {\bf (a,b)} Electron density distribution of electronic states in $n=4$ bowtie antennas for $m=0$ ((a), separated triangles) and $m=6$ ((b), bridged triangles). For each state of energy $E>0$, there is a state of energy $-E$ (not shown) with identical density distribution. Only low-order states close to the Dirac point (zero energy) are shown. The energies of these states are indicated by black lines right under the corresponding density plots. {\bf (c)} Dipole matrix elements between electronic states of the same bowtie as in (b). The area of the circles is proportional to the dipole strength. We only show transitions between states separated by energy differences larger than $|E_i-E_j|>0.01\,$eV. The solid lines connect initial and final energies corresponding to 0.47\,eV (BDP), 0.35\,eV (JP), and 0.22\,eV (CTP) excitations.} \label{Fig3}
\end{center}
\end{figure*}

The electronic states involved in the plasmon of separated nanotriangles [Fig.\ \ref{Fig3}(a)] are nearly identical to those of individual triangular islands, but with a small amount of Coulombic interaction and hybridization. When a bridge of intermediate width is added to form a junction [Fig.\ \ref{Fig3}(b)], the level of hybridization of these states increases, and new electronic junction states emerge. In particular, two new junction states are observed near the Dirac point (zero energy), which are expected due to the presence of carbon zigzag edges in the bridge \cite{AB08_2,WAG10}. Thus, the junction plasmon noted above must be supported by excitations involving electrons or holes in these electronic junction states. The strength of the electron-hole pair (e-h) dipole transitions in this system are represented in Fig.\ \ref{Fig3}(c) as a function of initial and final energies (the strength is represented through the area of the circles). Clearly, the plasmon energies (solid curves) do not strongly overlap with the dominant e-h transitions. This energy mismatch shows that the optical transitions are not single-particle excitations and is a strong argument supporting our conclusion of collective plasmonic nature of the bowtie optical excitations under consideration. Similar results are observed for different widths of the junction (see Appendix, Fig.\ \ref{states}).

\section{Outlook and conclusions}

Graphene dimers provide advantageous systems compared to metallic dimers both because of the structural robustness of this material and because the carbon structure can be reliably imaged with atomic detail \cite{GME09,SLL12}, and therefore, a detailed correlation between atomic structure and optical response is within reach experimentally. This is not the case of gaps formed by closely spaced three-dimentional metallic particles or tips \cite{KBK09,paper167}, which suffer from inherent faceting and structural uncertainties. Planar graphene nanostructures are thus ideal systems on which to study the interplay between structural features (e.g., ripples and defects) and the plasmonic response (for example, the amount by which they increase the plasmon scattering and dephasing rate via electron-hole pair generation). Recent advances in atomically resolved imaging \cite{TCW11} and patterning \cite{BFG12} of graphene should facilitate this task.

The remarkably large variations here reported in the optical properties of plasmonic structures consisting of thousands of carbon atoms by just adding a few ($<10$) extra atoms provides a handle for tuning plasmons in graphene. Together with the already established chemical \cite{WNM08} and electrical \cite{NGM04} tunability of this material, this adds a novel tuning parameter, which can be potentially exploited for ultrasensitive optical sensing (e.g., by analyzing the dependence on the presence of molecules near the junction, which can modify the local valence charge density).

Although for simplicity we only discuss above free-standing graphene with a Fermi energy $E_F=0.4\,$eV, no qualitative changes are observed for different doping or when the graphene sheet is placed on a substrate (see Appendix, Figs.\ \ref{SIEF} and \ref{substrate}). Quantum effects are found to be more significant when the level of doping is reduced (see results for $E_F=0.2\,$eV in the Appendix, Figs.\ \ref{SIEF}), as one expects for the smaller number of doping electrons participating in the plasmons, whereas the CTP energies are closer to the classical description when the doping increases (see $E_F=0.8\,$eV results in the Appendix, Figs.\ \ref{SIEF}).

In conclusion, using a fully quantum mechanical approach we have studied the optical properties of graphene dimers consisting of several thousand carbon atoms connected by a narrow graphene junction. The optical properties of the system are found to be strongly sensitive to the detailed structure of the junction, with the spectrum exhibiting drastic changes with the addition of a few carbon atoms in that region. In addition to the conventional hybridized and charge transfer plasmons observed for metallic dimers \cite{paper075,LAH08,ZPN09,EBN12}, graphene dimers display junction plasmons of a purely quantum mechanical origin, which does not seem to appear in conventional metallic junctions \cite{SHE12}. The plasmonic response of this graphene structure is found to be strongly influenced by quantum effects, making it of significant interest in the emerging field of quantum plasmonics as well as for novel nanophotonics and optoelectronics applications.

\section*{ACKNOWLEDGMENT}

This work has been supported in part by the Spanish MEC (MAT2010-14885 and Consolider NanoLight.es) and the European Commission (FP7-ICT-2009-4-248909-LIMA and FP7-ICT-2009-4-248855-N4E). A.M. acknowledges financial support through FPU from the Spanish MEC.  P.N. acknowledges support from the Robert A. Welch Foundation under grant C-1222.

\section*{APPENDIX}

In this appendix, we provide detailed results of first-principles extinction spectra, electron states, and plasmon-induced charge densities for the graphene bowtie structures under consideration. Additionally, we study the evolution of plasmons in bowties of increasing length. Classical-electrodynamics calculations are also offered for comparison.

\begin{figure*}
\begin{center}
\includegraphics[width=100mm,angle=0,clip]{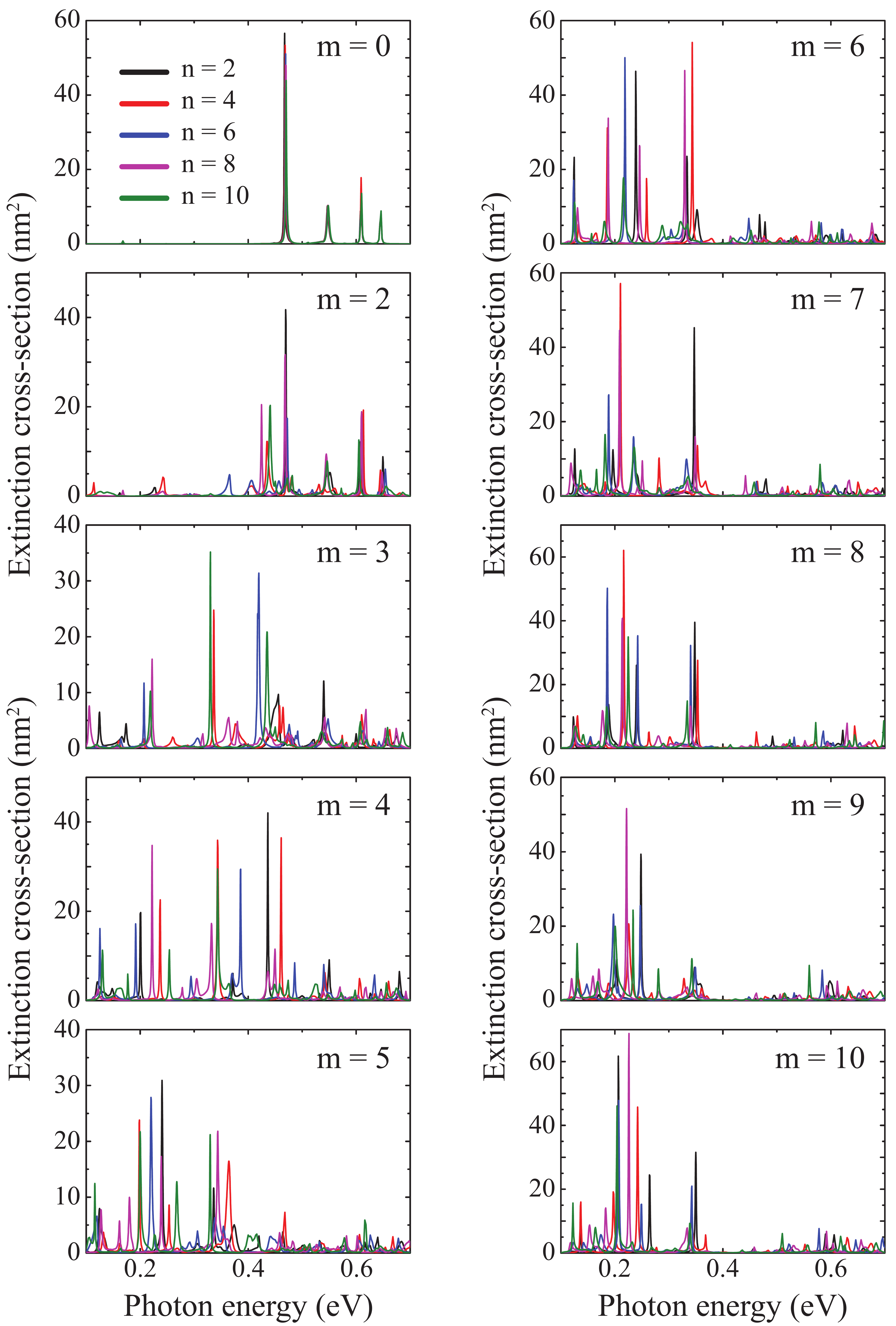}
\caption{{\bf Extinction spectra in bowties of increasing junction width.} We here show the full extinction spectra from which Figs.\ \ref{Fig1}(e) and \ref{Fig1}(f) are constructed as a function of junction width $m$ and length $n$.} \label{ext}
\end{center}
\end{figure*}

\begin{figure*}
\begin{center}
\includegraphics[width=160mm,angle=0,clip]{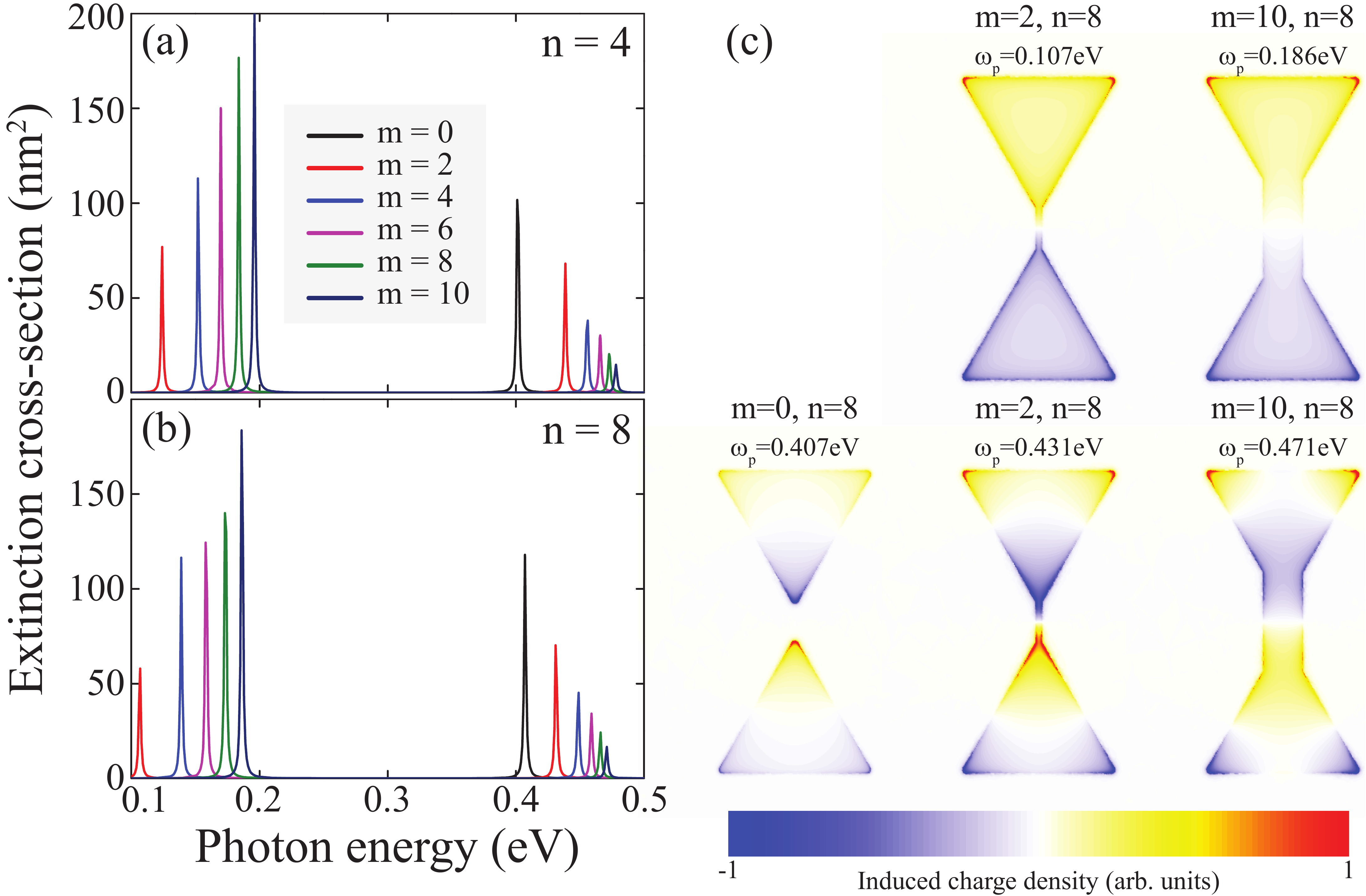}
\caption{{\bf Classical-electrodynamic calculations.} {\bf (a,b)} Extinction spectra of bowties for two different values of the junction length $n$ (we use the same notation as in Sec.\ \ref{results} for $m$ and $n$). {\bf (c)} Induced charge distributions associated with representative plasmons in these structures. Upper row: charge-transfer plasmons. Lower row: dipole-dipole hybridized plasmons.} \label{classical}
\end{center}
\end{figure*}

\begin{figure*}
\begin{center}
\includegraphics[width=110mm,angle=0,clip]{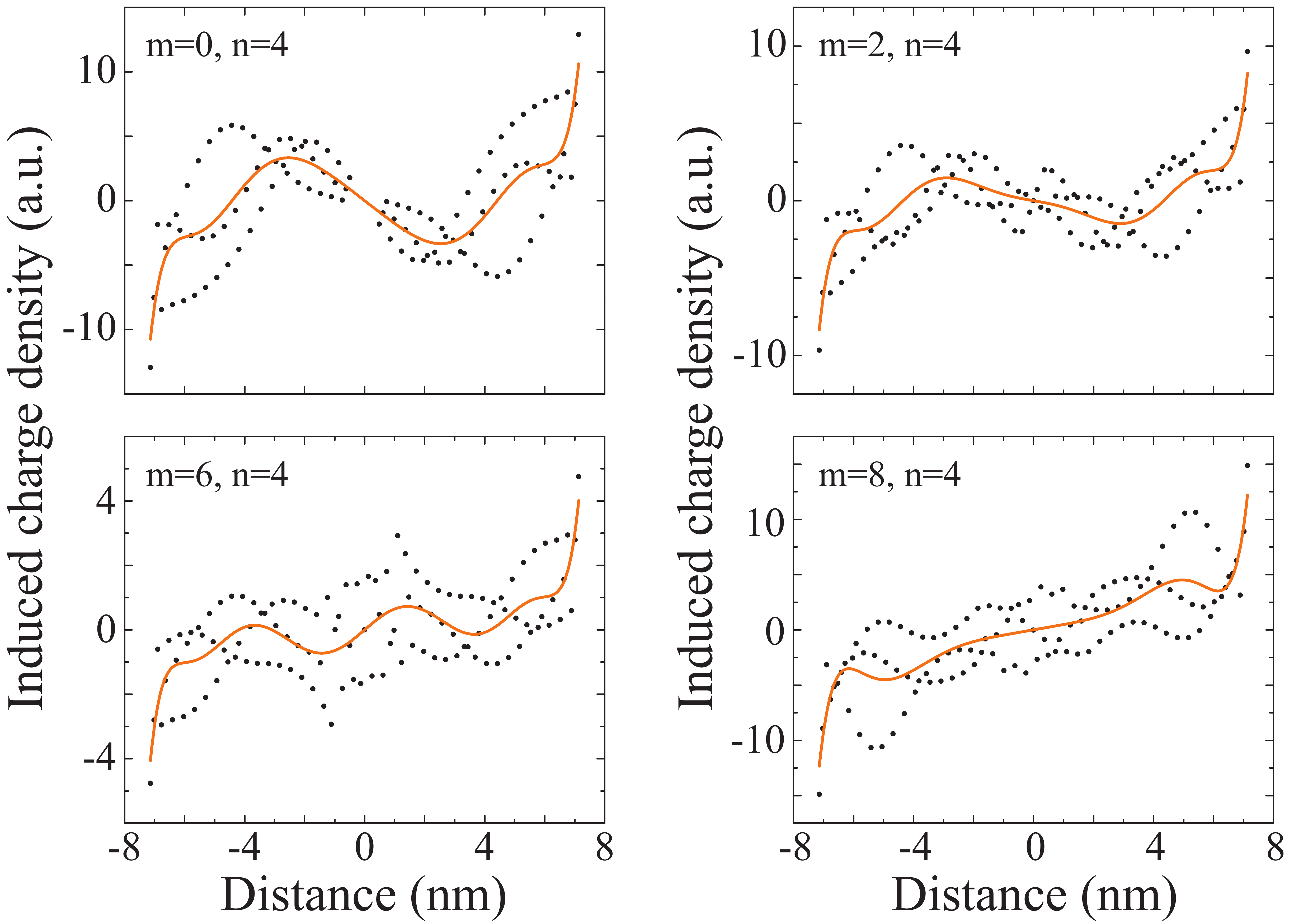}
\caption{{\bf Plasmon charge density along the bowties.} Induced charge associated with the plasmons considered in Fig.\ \ref{Fig2}. Dots: Charge density integrated across the width of the bowtie as a function of the position of the carbon atoms relative to the center of the junction. Curves: Same, averaged over four nearest-neighbor atom positions along the axis of the bowtie.} \label{charge}
\end{center}
\end{figure*}

\begin{figure*}
\begin{center}
\includegraphics[width=110mm,angle=0,clip]{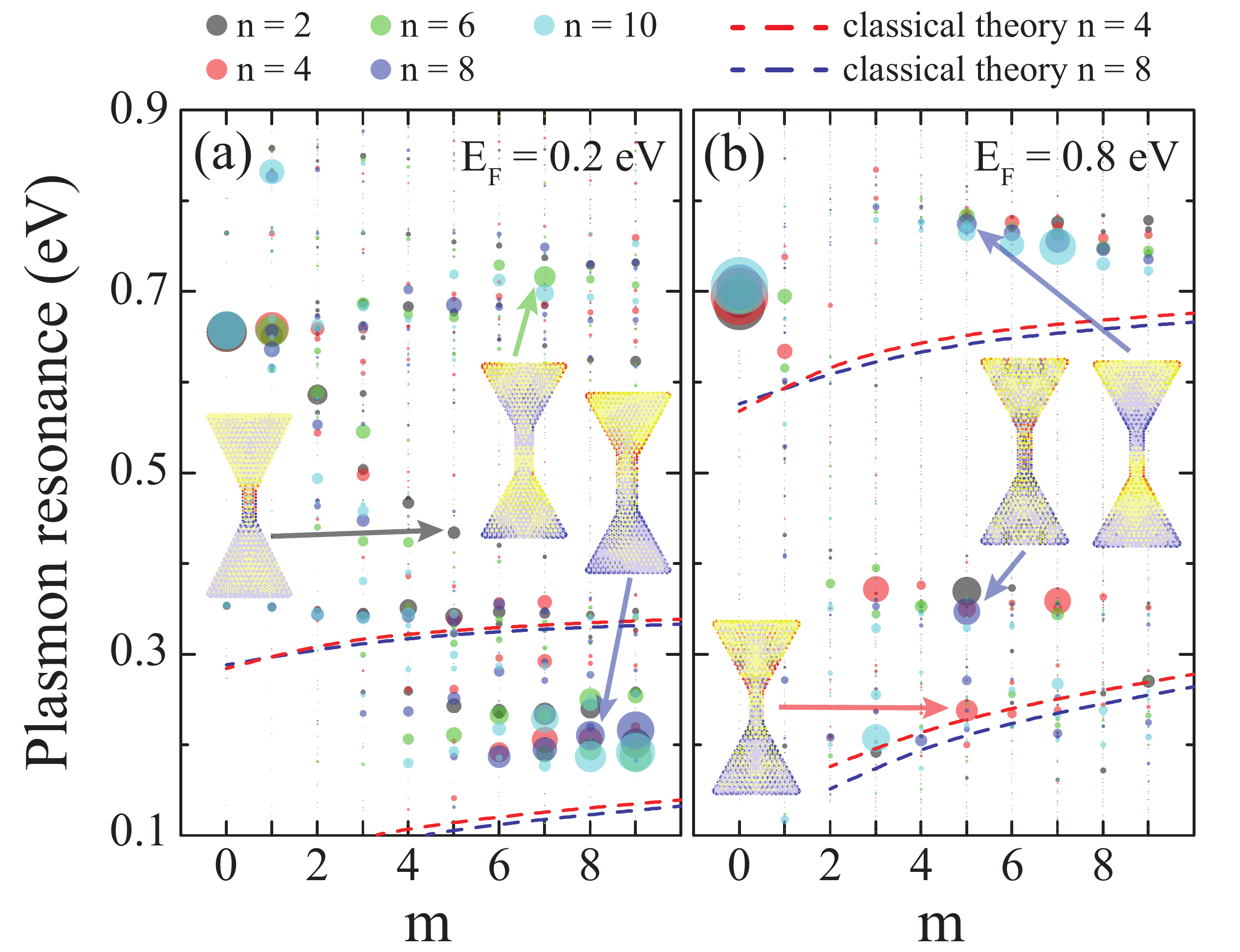}
\caption{{\bf Effect of doping.} We show the plasmon energies and strengths for two different levels of doping ($E_F=0.2\,$eV in (a) and $E_F=0.8\,$eV in (b)) with the same notation and style as in Fig.\ \ref{Fig1}(e). The insets show plasmon charge distributions for selected geometries.} \label{SIEF}
\end{center}
\end{figure*}

\begin{figure}
\begin{center}
\includegraphics[width=60mm,angle=0,clip]{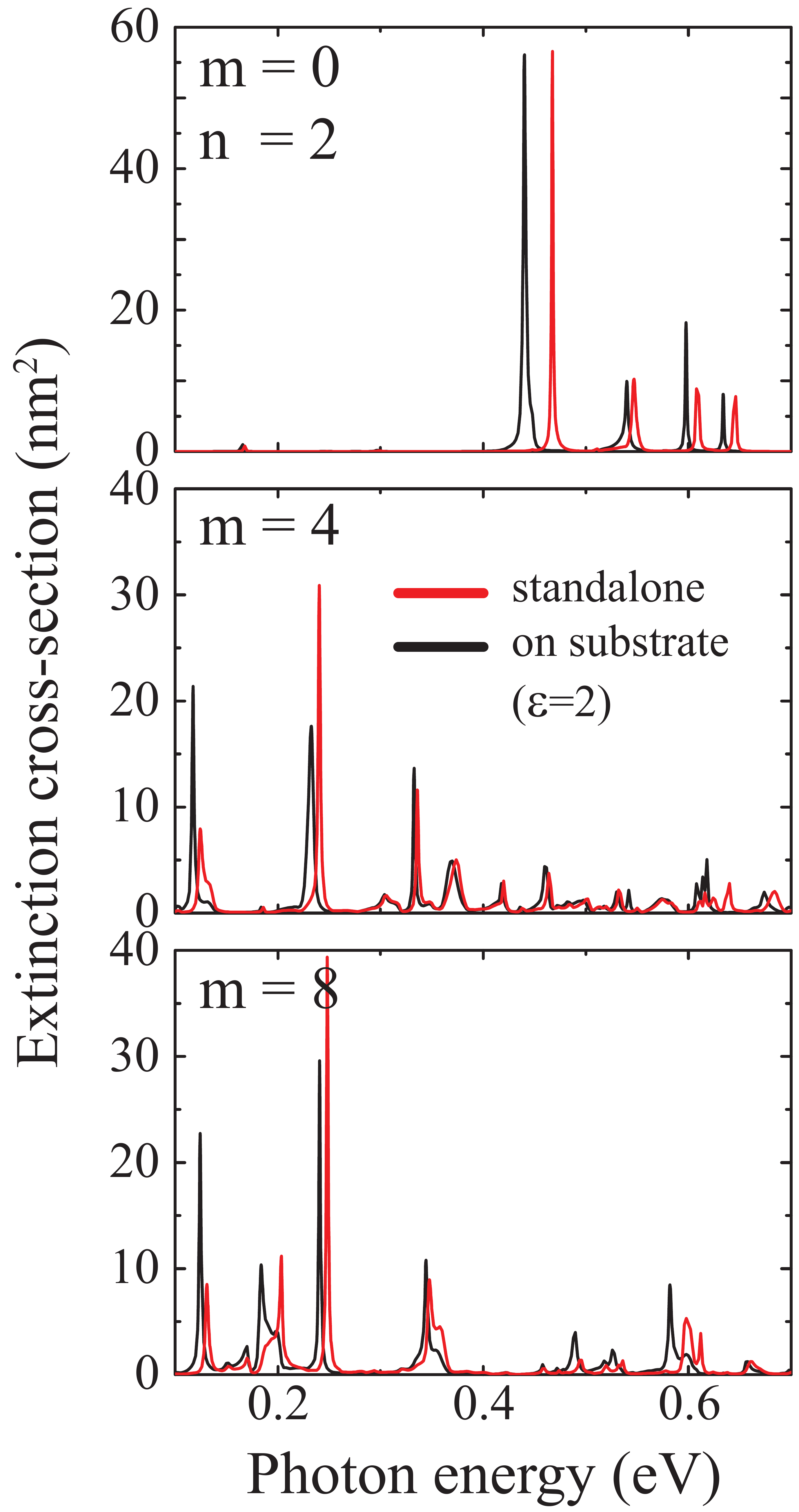}
\caption{{\bf Effect of the substrate.} Extinction spectra of self-standing graphene bowties (red curves) and graphene supported on glass (black curves) for different values of the junction width $m$ and fixed length $n=2$.} \label{substrate}
\end{center}
\end{figure}

\begin{figure*}
\begin{center}
\includegraphics[width=180mm,angle=0,clip]{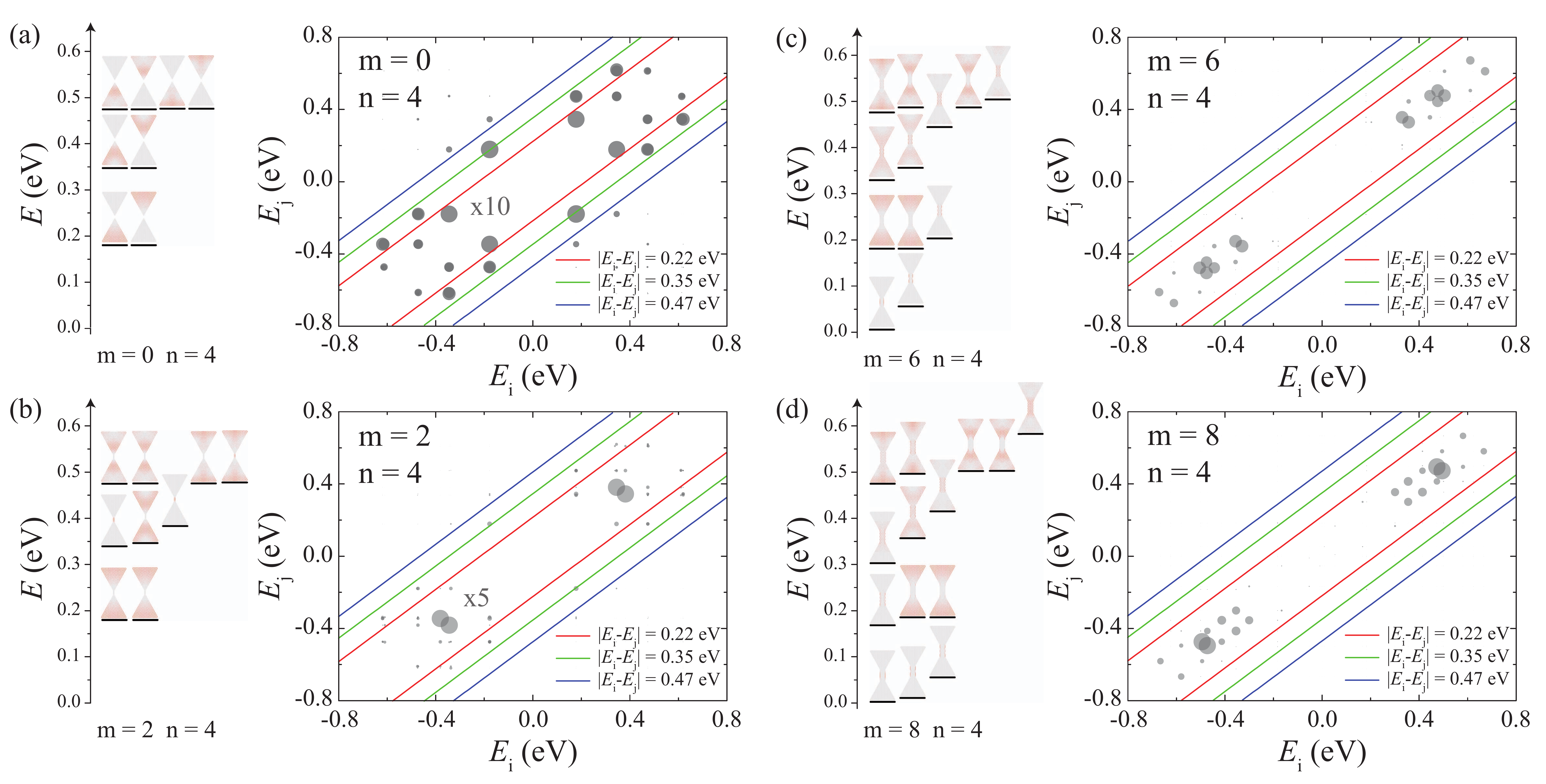}
\caption{{\bf Electron states and electron-hole transition strengths.} We show here a more extended version of Fig.\ \ref{Fig3}, with electron states and electron-hole transition strengths plotted for various values of the junction width $m$ and length $n$. We only show transitions between states separated by energy differences larger than $|E_i-E_j|>0.01\,$eV.} \label{states}
\end{center}
\end{figure*}

\begin{figure*}
\begin{center}
\includegraphics[width=177mm,angle=0,clip]{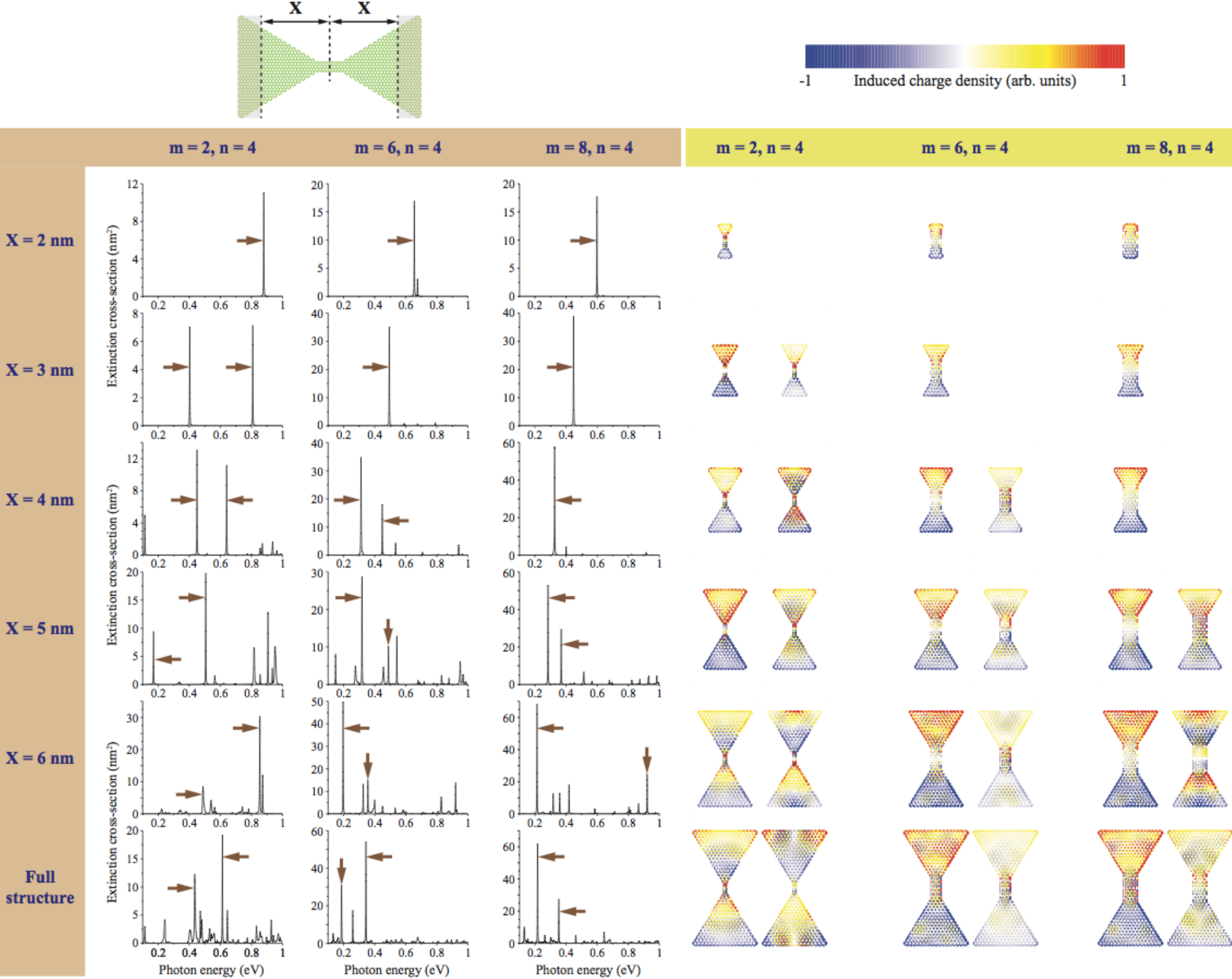}
\caption{{\bf Formation of different types of plasmons in bowtie antennas of increasing length.} Left: extinction spectra for three different values of the junction width $m$ and increasingly larger bowties. The half-length of the bowtie is denoted $X$ (see upper inset). The full structure corresponds to a triangle side-length of 8\,nm (i.e., $X\approx7\,$nm). Right: plasmon charge density distributions for the features indicated by arrows in the extinction spectra (in the same order from left to right within each row).} \label{cut_end}
\end{center}
\end{figure*}

We choose a Fermi energy $E_F=0.4\,$eV (except in Fig.\ \ref{SIEF}) and an intrinsic damping $\hbar\tau^{-1}=1.6\,$meV, corresponding to a DC mobility of 10,000\,cm$^2/($Vs$)$. We set the temperature to $T=300\,$K, although thermal effects are negligible for the plasmons under consideration, the energies of which are high compared with $k_BT$. The external light polarization is taken along the junction between the nanotriangles forming the bowties. The triangle side-length is set to 8\,nm, except in Fig.\ \ref{cut_end}. Our first-principles calculations are performed following methods described elsewhere \cite{paper183}. For the classical calculations of Fig.\ \ref{classical} we use a finite-element method (COMSOL) following a previously reported prescription \cite{paper176} according to which the graphene is treated as a thin dielectric layer.

We show in Fig.\ \ref{ext} the full series of extinction spectra from which Figs.\ \ref{Fig1}(e) and \ref{Fig1}(f) are constructed. The cross section is given in square nanometers, so it can be directly compared with the geometrical area of the graphene in the bowties $\approx111\,$nm$^2$. The maximum cross-section obtained in these first-principles calculations is approximately half the geometrical area. In contrast, the cross sections derived from classical electrodynamics (Fig.\ \ref{classical}) can reach substantially large values compared with the geometrical area. The energies of the plasmons derived from classical calculations are represented in Fig.\ \ref{Fig1}(e) for comparison.

The charge densities associated with the plasmons (Fig.\ \ref{Fig2}) are examined in more detail in Fig.\ \ref{charge}, where they are shown to exhibit strong modulations between nearest-neighbor carbon-atom rows. These modulations oscillate between three smoothly varying limits, as expected from the three different types of atomic sites in the armchair edges of the nanotriangles under consideration. The curves plotted in Fig.\ \ref{Fig2}, and reproduced in Fig.\ \ref{charge} for convenience, correspond to averages over four carbon-atom neighbors.

We have considered the effect of doping in Fig.\ \ref{SIEF}, where we plot data similar to Fig.\ \ref{Fig1}(e) ($E_F=0.4\,$eV), but changing the doping level to $E_F=0.2\,$eV (Fig.\ \ref{SIEF}(a)) and $E_F=0.8\,$eV (Fig.\ \ref{SIEF}(b)). As a general trend, we observe a reduction in the role of quantum effects as the doping level is increased. For example, for $E_F=0.2\,$eV we observe that charge-transfer plasmons (CTPs) pile up around $0.15\,$eV in the quantum description, whereas classical theory places them at much lower energies; in contrast, both classical and quantum descriptions agree much better in the energies of CTPs for $E_F=0.8\,$eV; and an intermediate situation is encountered for $E_F=0.4\,$eV. Likewise, junction plasmons (JPs) appear to lose strength relative to the rest of the features as $E_F$ increases. Actually, these plasmons cover a wide spectral region extending above the energies of CTPs for $E_F=0.2\,$eV as the junction width is varied, whereas they are only discernable in the $m=3-7$ range of junction widths for $E_F=0.8\,$eV.

For simplicity, we present in this paper results for self-standing graphene, except in Fig.\ \ref{substrate}, where we study the effect of depositing the carbon layer on a glass substrate ($\epsilon=2$). In a classical description of graphene, the net effect of the substrate is to change the conductivity $\sigma$ to an effective value $2\sigma/(\epsilon+1)$ \cite{paper181}. This leads to a redshift of the plasmon energies by a factor $\sim\sqrt{2/(\epsilon+1)}$. We show in Fig.\ \ref{substrate} that a similar trend is observed in small graphene structures within our quantum-mechanical description, thus corroborating that the conclusions of our study also apply to the more physical situation of supported graphene nanostructures. The features of the extinction spectra are maintained by introducing a substrate, which essentially produces a redshift that is substantially smaller than that predicted by classical theory in our relatively small structures, particularly for the lowest-energy features and the widest junction considered in Fig.\ \ref{substrate}.

The electron wave functions and the strength of dipolar transitions connecting them are considered in Fig.\ \ref{states}, which is an extension of Fig.\ \ref{Fig3} including more junction widths. One can clearly observe the increasing involvement of electron junction states as the junction becomes wider.

Finally, we study in Fig.\ \ref{cut_end} the evolution of plasmons in bowties as a function of the length of the structures. We fix the junction length to $n=4$ hexagons and consider three different junction widths consisting of $m=2$, 6, and 8 carbon zigzag rows. In general, we observe a migration of modes with similar symmetry towards lower plasmon energies as the bowtie length increases. For the two shorter series under consideration (half length $=2$, 3, and 4\,nm), we find a dominant CTP, with opposite induced charges in each half of the dimer. For larger bowties, higher-order modes become relevant. For example, the structure with the narrower junction ($m=2$) exhibits a dipole-dipole plasmon, which ends up at $\approx0.4\,$eV at the full length $2X\approx14\,$nm (lower row of Fig.\ \ref{cut_end}). For this junction width, the CTP becomes relatively weak with $X\ge6\,$nm, so it does not make a relevant contribution to the full structure. In contrast, the CTP remains a dominant feature for all the lengths under consideration in structures bridged by wider junctions ($m=6$ and 8). As the length increases, a new dipole mode starts developing, which is associated with large charge pileup at the junction. This is the intermediate-energy feature observed in Fig.\ \ref{Fig1}(e).


\end{document}